# Methodology and Analysis of Smart Contracts in Blockchain-Based International Trade Application

Asif Bhat[1][0000-0001-5695-520X], Rizal Mohd Nor[1][0000-0002-8994-2234], Md. Amiruzzaman[2][0000-0002-2292-5798] and Md. Rajibul Islam[3][0000-0003-0565-6917]

[1] Kulliyah of Information and Communication Technology International Islamic University Malaysia Kuala Lumpur, Malaysia
bhatasif.17@gmail.com, rizalmohdnor@iium.edu.my
[2] West Chester University, PA, USA
mamiruzzaman@wcupa.edu
[3] Department of Computer Science and Engineering University of Asia Pacific
md.rajibul.islam@uap-bd.edu

**Abstract.** Blockchain is used in a variety of applications where trustworthy computing is required. Trade finance is one of these areas that would benefit immensely from a decentralized way of doing transactions. This paper presents the preliminary assessment of Accepire-BT, a software platform developed for the practice of collaborative Trade Finance. The proposed solution is enforced by smart contracts using Solidity, the underlying programming language for the Ethereum blockchain. We evaluated the performance in the Rinkeby test network by using Remix and MetaMask. The results of the preliminary trial show that smart contracts take less than one minute per cycle. Also, we present a discussion about costs for using the public Ethereum Rinkeby network.

## 1 Introduction

Blockchain technology has been widely debated and in recent articles it explores the position blockchain plays in business operations, cultural aspects, and other fields [1-3], such as in international trade. In this paper, we study international trade, its business processes between different types of trading parties as well as paper-based market activities. Paper-based market activities involve information exchanges, asset exchanges, handovers of goods, and a payment workflow to facilitate trade finance. Historically, business parties obtain trust and legitimacy from a centralized system for payment by Letter of Credit (L/C). However, system speed and processing performance and susceptibility to malicious alterations are significant detriments.

As one of the most powerful methods for ensuring the continuity of the reproductive process, introducing different creative strategies in trade finance will solve current economic problems. Thus, often trade finance's inaccessibility to replenish working capital among small and medium-sized enterprises is acutely felt. One of the few reasons are due to insufficient useful resources to provide clarity and reduce asymmetric knowledge issues. Furthermore, the development of trade finance has several other difficulties, firstly the volume of transactions for trade financing is dynamic. External factors, such as interest rates, regulatory changes, and internal factors, such as limited capital and the need for depreciation, have an impact on system development [2]. Secondly, the introduction of trade financing in the global supply chain is challenging. Implementation requires tremendous vendor efforts and careful integration with finance, purchasing, and IT divisions within the company. Thirdly, it is difficult for various participants to have incentives. In terms of asymmetrical systems, costs, and uncertainties, suppliers, consumers, and financiers are autonomous decision-makers who aim to optimize income. These independent maximizations of profit also lead to low output in the entire supply chain.

In this paper, we explore our experience gained from applying the blockchain to a prototype framework for trade finance. The prototype was implemented on the Ethereum blockchain. This paper's essential contribution is the methodology used and study of smart contracts in trade finance based on blockchain and describe how business processes can be simplified by blockchain and related smart contract technologies. It also discusses the roles of smart contracts' in streamlining trade finance operations and the value they could provide in terms of re-designing processes. We used the test network (Rinkeby), which is currently one of the most common blockchains implementing smart contracts, and it helps to evaluate performance. We also assess the financial costs associated with the environment. Additionally, we structured our research approach with the Design Science Research (DSR) framework similar to work in Hevner et al. [6] in which the core concept of design science is the artifact: an object that can be instantiated with physical or social properties.

This paper proceeds with the following sections, Section II describes the literature review of smart contracts, Section II explains the Methodology used followed by Experiments accomplished in Section IV. Section V describes the Data and Preliminary Results. Section VI concludes the paper.

## 2      Literature Review

The word smart contract came long before Bitcoin and blockchain. A smart contract was described by Szabo (1994) [4] as a computerized transaction protocol that satisfies contractual conditions in terms of payment, confidentiality, or compliance, reduces exceptions, and minimizes the need for trusted intermediaries. As an example of a smart contracts, he discussed digital cash protocols a mechanism to allow online payments with paper cash characteristics while considering divisibility and confidentiality. Szabo (1997) [5] later defined smart contracts as "the combination of protocols with user interfaces to ensure structured and secure network connections". The legal, economic, and technological foundations are the basis for the design of such structures. Smart contracts, therefore, involve interdisciplinary research. Smart contracts quality and their usability in legally binding contracts may be differentiated by multitude of degrees depending on:
- A necessary computer code that does not constitute a legal agreement but merely implements a predefined logic.
- Computer code that has specific legal properties, i.e., a program based on legal structures with a predefined logic that is supposed to behave in a certain way or
- The (partial) execution by computer code of a legal person (e.g., a contract) where the code resembles the legal person.

A significant factor and development goals in resolving these problems in trade finance instruments is to improve software and utilization of blockchain technologies. These blockchain technologies allow companies, through partnerships and process automation throughout the supply chain, to unite and accelerate cash flow and documentation within the supply chain [2].

Decentralization of the operating business networks will improve issues such as accountability, real-time monitoring, and trustless player transactions [3]. Additionally, PingNET IOT devices will improve Accepire-BT to provide transparency of transactions and traceability of the supply chain. Smart contracts are implemented to execute event-based contract terms or agreements such as Sales Contract, Financial Contract, Letter of Credit Contract, Shipment Initiated Contract, Shipment Received Contract, Payment Contract, etc. which are important to execute a Trade Finance transaction. The following paragraphs present the use of blockchain in different types of transactions.

An example of artifacts can be diverse such as software, models, or norms. In [6], Hevner et al. conceptualized design science research within information systems (IS) research with the following three integrated dimensions:
- The environment including people, organizations, and technology,
- IS research pinpointing the creation and justification of artifacts, and
- The knowledge base bringing forward foundations and methodologies to be used in the creation and evaluation of artifacts.

## 3   Methodology

For this research, we will be utilizing Hevner et al. (2007)'s three-cycle view. This explains the steps for development and emphasizes evaluating the artifact and the contributions to the knowledge base and environment. We see this as the most readily applicable model and the most relevant for developing our prototype for our research. As highlighted in the research method by Hevner (2007), it is essential to show its novelty [7]. The framework consists of three significant development stages, Relevance, Rigor, and the Design cycle, as shown in Fig I. Each of these stages consists of multiple elements. In fig. 1, it is shown how we will be applying the model. This is demonstrated through the various stages within each of the cycles in the model.

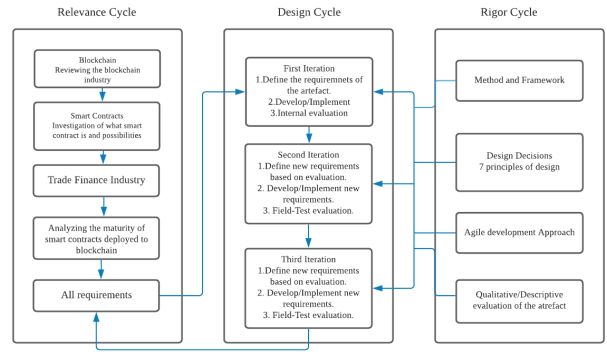

**Fig. 1.** Overview of the three cycles applied in this research

1. Relevance Cycle

The focus of the relevance cycle is two-part. In the first part, we establish a problem and determine the elements required to create the solution. The relevance cycle of our study consists of the three elements of our literature analysis, a review of blockchain technologies, smart contracts, and a review of the current business model of the trade finance industry. The second part of our relevance cycle is determining the requirements for building the artifact. This step consists of two parts. First, we gather knowledge of how deployment of smart contracts on the blockchain works and then by combining the different elements in the relevance cycle.

2. Design Cycle

In the design cycle, we develop the artifact using the frameworks from the rigor cycle. We compare the artifact to the relevance cycle requirements. After each development phase, we apply the evaluation method introduced in the rigor cycle to evaluate and to establish requirements for the next iteration.

3. Rigor Cycle

This study's rigor cycle consists of the methodological framework, the design decisions, and the artifact's qualitative and descriptive evaluation. This research utilizes design science research as a methodological framework, and as described uses, the three-cycle design view for the development of the artifact. The rigor contributes to the framework as well as the design decisions for the development cycle. In this phase, we introduce the evaluation methods for the design cycle. These evaluations will serve as the basis for the next iteration.

### 3.1 Data Gathering

The data gathering in our research consists primarily of qualitative data research, assessing methodological research and frameworks. To develop a prototype, the preferred framework three-view design cycle [1] is used. The theory used for this study is qualitatively gathered and limited mainly to a few sources because the technology is new and innovative. The achieved data gathering is through extensive research on articles, webpages, and books revolving around blockchain technology.

Through each iteration of the design cycle, it will be an evaluation of the current design. Chang et al. (2019) also highlights this as an essential and significant part of design science research [1]. We will be evaluating the prototype throughout the design cycle in various ways.

In this study, we will have three iterations. The authors will internally review the first iteration using functional black box testing, which is mentioned by Hevner et al., 2004 [6]. We will be discussing the potentials of blockchain technology, which will be approached in the first development phase.

The second and third iterations evaluation phase is conducted through a constructed scenario evaluation approach [8]. The second evaluation phase is the first field-tested iteration. We will conduct in-depth interviews with people familiar with the trade financing industry. The third iteration will develop on the key learnings from the second iteration and be re-evaluated by both evaluators from the second iteration and new evaluators. We will be doing this to ensure that we have both new eyes on the artifact and confirm by some users from the previous iteration that the latest design solution has improved the previous evaluation issues. The test users are found through our network of people who have experienced trade financing before testing our prototype. This should result in further ideas and suggestions, which we will then be evaluating.

The blockchain provides improvements in various aspects of trade finance. In this study, we focused on six major dimensions of the blockchain: (i) transparency, (ii) information transmission, (iii) traceability, (iv) disintermediation, (v) cost, and (vi) the incorporation of IoT to analyze the impact of blockchain-based trade finance, particularly in supporting an L/C payment process.

Cost is one of the critical concerns of supporting processes for L/C finance. In cross-border transactions, businesses have invested substantial sums in minimizing trade-related administrative data. In other payment options, such as inter-firm trade credit, handovers across shipping routes have decreased process flow efficiency and made business less competitive. As opposed to conventional payment systems, the blockchain will streamline trade processes and demand less.

The fundamental explanation is that the cost of centralized service providers is reduced. Significant savings in the sense of global trade are compensated for by the removal of paperwork. The blockchain ensures substantial cost savings with its immutable shared ledger. Manual paper-based processes are also significantly reduced, eliminating the time and workforce necessary for processing documents. The case of Maersk gave practitioners faith in the way digitized properties and documents are moved across borders [1]. Blockchain implementations often exploit their ability to reduce associated costs, such as those related to search, negotiation, and policing, from the perspective of transaction cost economics.

In the following section, we measure the migration costs and transaction fees for each of the smart contracts that are to be utilized in a trade finance transaction.

## 4     Experiment

In developing our Ethereum smart contracts build in Solidity language, the following technology choices are taken to deploy our smart contract. This section describes the steps taken to deploy our smart contracts designed for trade finance.

### 4.1     Environment Setup

The Remix is a web-based Integrated Development Environment (IDE) for creating, running, and debugging smart contracts in the browser. It is developed and maintained by the Ethereum foundation. Remix allows Solidity developers to write smart contracts without a development machine since everything required is included in the web interface. It allows for a simplified method of interacting with deployed contracts without the need for a command-line interface.

Remix encourages a fast development cycle and has a rich collection of Graphical User Interface (GUI) plugins. The Remix is used in this research for the whole contract development journey. It is a powerful open-source platform to help write solidity contracts from the browser directly. It is written in JavaScript and supports both usages in the browser, in the browser but run locally and the desktop edition. Remix IDE has modules to test, debug, and deploy smart contracts and much more.

The language used to write smart contracts is called Solidity. Solidity is a high-level, object-oriented language for smart contracts implementation. Smart contracts are programs that regulate the conduct of accounts in the Ethereum State. Solidity has been influenced by C++, Python, and JavaScript and is targeted at Ethereum Virtual Machine (EVM). Solidity is statically typed, supports inheritance, libraries, and complex user-defined types, among other features. For this paper, Solidity version 0.7.5 is used.

For file storage, we took a decentralized approached which means there should not be any local databases or distributed data centers. This is where the IPFS platform comes into the picture. By storing the files on IPFS, the file storing aspect of our artifact can be considered decentralized. Of course, some technical considerations must be made before implementing this, such as ensuring that only the correct people can read the files. This could be done via encryption. The IPFS is a protocol and network designed to provide a peer-to-peer storage and sharing mechanism for content-addressable in a distributed file system. Like Torrent, IPFS not only allows users to receive but also to host content. Compared to a centrally located server, a decentralized system is designed for the user operators who hold a fraction of the overall data to build and distribute a resilient file storage system. In other words, a high-performance content-addressed block storage model is provided by IPFS with content hyperlinks. This is a simplified Merkle DAG, a data structure where versioned file systems, blockchains, and a permanent web can be built. IPFS contains a hash table, a block exchange, and a namespace to authenticate itself. There is no single point of failure and nodes does not need to trust each other [9]. IPFS is decentralized much like blockchain and the security is provided by the IPFS "link" that generates uniquely every time a document is updated or changed.

In this paper, MetaMask is used to test the smart contracts. MetaMask was developed to meet the needs of secure and accessible websites based on Ethereum. It controls account management and connects the user to the blockchain. It integrates with Remix and is used to test smart contracts. Should there be a need for EPR-systems to communicate via short messages, the Whisper protocol is then utilized.

### 4.2 Deploying Smart Contracts

In our experiment, we created 3 smart contracts, sales, financial and letter of credit shown in fig. 2 below.

| ⓘ To: | [Contract 0x65663423b9886185aa |
|---|---|
| ⓘ Value: | 0 Ether ($0.00) |
| ⓘ Transaction Fee: | 0.000440383 Ether ($0.000000) |
| ⓘ To: | [Contract 0x65663423b9886185aa |
| ⓘ Value: | 0 Ether ($0.00) |
| ⓘ Transaction Fee: | 0.000440383 Ether ($0.000000) |
| ⓘ To: | [Contract 0xe9bcf98413efca495ba |
| ⓘ Value: | 0 Ether ($0.00) |
| ⓘ Transaction Fee: | 0.000640725 Ether ($0.000000) |

**Fig. 2.** The Sales, Financial and Letter of Credit Smart Contract deployed on the Rinkeby Ethereum Test Network with corresponding transaction fees.

Before a smart contract instance can be invoked, it needs to be deployed on the network. Deploying these contracts requires a one-time deployment cost. The total migration cost for all three contracts is 0.00138554 + 0.000440383 + 0.0006407250 = 0.002466648 ETH, which is about 1.358 USD (1 ETH = 550.75 USD).

## 5 Results

To determine the cost of each implementation, each method's transaction fee was calculated using the Remix, MetaMask, and Rinkeby Test Network in the Ethereum Solidity Code. When executing a function, the Remix console logs the transaction cost.

## 5.1 Gas Cost

At the time of this research, the gas price was 0.000000001ETH (1ETH=550.75USD), and the transactions per second is 15 TPS. The formula for calculating transaction fee is given in equation (1) below where $t_c$ = transaction cost, $n$ = number of characters, $g_c$ = gas per character > 0, $g_p$ (gas price) = 0.000000001 ETH and $t_f$ = transaction fee.

$$(t_c + (n * g_c)) * g_p = t_f \qquad (1)$$

**Table 1.** Contract, functions, and associated cost.

| Contract | function | Trans Cost | Gas Price (ETH) | Trans Fee (ETH) | * Cost (USD) |
|---|---|---|---|---|---|
| Sales | setSalesContract | 106384 | 1E-09 | 0.00010638 | 0.06 |
| | **addOrder | 176983 | 1E-09 | 0.00017698 | 0.10 |
| | createInvoice | 109016 | 1E-09 | 0.00010902 | 0.06 |
| | confirmInvoice | 43758 | 1E-09 | 4.3758E-05 | 0.02 |
| | confirmOrder | 47653 | 1E-09 | 4.7653E-05 | 0.03 |
| | orderExists | 0 | 1E-09 | 0 | 0.00 |
| | cancelOrder | 45495 | 1E-09 | 4.5495E-05 | 0.03 |
| | receiveOrder | 43734 | 1E-09 | 4.3734E-05 | 0.02 |
| Financial | setFinancialAgreementParties | 127510 | 1E-09 | 0.00012751 | 0.07 |
| | confirmAgreement | 44678 | 1E-09 | 4.4678E-05 | 0.02 |
| Letter of Credit | initializeContract | 169459 | 1E-09 | 0.00016946 | 0.06 |
| | addDocument | 68518 | 1E-09 | 0.000177 | 0.10 |
| | getNumberOfDocuments | 0 | 1E-09 | 0 | 0.00 |
| | getDocumentID | 0 | 1E-09 | 0 | 0.00 |
| | IsDocumentValid | 0 | 1E-09 | 0 | 0.00 |
| | validateDocument | 45242 | 1E-09 | 4.5242E-05 | 0.02 |

In our observation, the cost for each function calls are collected and shown in table 1. It can be summarized to the nearest decimal point, that the cost for total calls for functions in sales is about 0.31 USD, financial functions is about 0.09 USD and for Letter of Credit functions is about 0.22 USD. The cost of some functions is not constant due to the use of variable-length strings. This is especially true in Table 1 function **addOrder as it depends on the size of an order. Some functions' cost is 0 since it is a function that does not process any data and only returns a value.

## 6 Discussion

Perhaps the biggest scalability problems with Ethereum are that all transactions must be processed by every node, and the entire state of every account balance, contract code, and storage must be stored, etc. Although this offers a large amount of security, scalability is dramatically restricted to the extent where a blockchain cannot handle more transactions than a single node.

A potential solution to this problem is to create a new system where a small subset of nodes must only check a subset of transactions. The system will still be safe as long as there are enough nodes to validate each transaction, but it will also allow the system to process transactions in parallel. That method is called sharding. By splitting the global state of accounts, both external and contract accounts, smaller chunks known as a shard are the fundamental concept behind sharding.

In more complex types of sharding, transactions can also affect other shards in certain instances and may also request data synchronously from the state of several shards. Each shard gets its own set of validators, and not all shards would need to be validated by these validators [8].

Using a modified version of the GHOST protocol, Ethereum mitigated most security losses with faster block time. However, the blocks still have to propagate across the network, which is a relatively slow operation. To propagate faster, the block size needs to be smaller, and that is why Ethereum can only process around 15 TPS, even though the block time is as low as 15 seconds.

Ethereum is Turing Complete, and the network needs to manage random processing tasks and probably store large quantities of data. That is why, compared with Bitcoin, Ethereum is working on the most promising scaling solutions [10].

Accepire-BT guarantees transparency, reliability, and traceability using IOT devices known as PingNet combined with a decentralized oracle known as ChainLink [11].

We argue that the smart contract alone will not solve all of the supply chain's problems but will contribute to ensuring transparency and preventing fraud in safe transactions. The technology has not yet reached its full potential, but research is underway to make it more effective.

## 7 Conclusion

In conclusion, this paper introduces a blockchain-enabled trading finance methodology and smart contract examination. The application of smart contracts to control trade finance allows for clear and automated actions that guide stakeholder engagement. Blockchain's trade finance can provide solutions relating to fraud, confidence, audibility, accountability, processing time, and cost. It can be claimed that the mechanism of

10conventional trade can be streamlined and optimized with the use of blockchain platforms such as Accepire-BT. It is capable enough to create trust between a network with the security features provided by blockchain technology. It maintains the integrity of the information shared and can assist in tracking the whole process. Many laws and requirements must be enforced to ensure the legitimacy of international trade. Using blockchain can ease international trade auditability and can have a big effect on simplifying the objective of protecting society and the economy. We conclude here by claiming that blockchain will bring about disruptive changes in trade and trade finance.

A proof-of-concept will be presented in future work, describing the implementation using Solidity of the blockchain-powered trade finance model in Ethereum. Besides, by incorporating practitioners into our research process, we can look to further validate the model. Finally, before they can be implemented into a live blockchain, the supplied smart contract must be optimized in transactional cost and network cost.

## References

1. S. E. Chang, H. L. Luo, and Y. Chen, "Blockchain-Enabled Trade Finance Innovation: A Potential Paradigm Shift on Using Letter of Credit," Sustainability, vol. 12, no. 1, p. 188, Dec. 2019, doi: 10.3390/su12010188.
2. A. v. Bogucharskov, I. E. Pokamestov, K. R. Adamova, and Z. N. Tropina, "Adoption of blockchain technology in trade finance process," Journal of Reviews on Global Economics, 2018, doi: 10.6000/1929-7092.2018.07.47.
3. Raluca Gh. Popescu and Popescu, "An Exploratory Study Based on a Questionnaire Concerning Green and Sustainable Finance, Corporate Social Responsibility, and Performance: Evidence from the Romanian Business Environment," Journal of Risk and Financial Management, 2019, doi: 10.3390/jrfm12040162.
4. N. Szabo, "Smart Contracts," 1994. [Online]. Available: https://www.fon.hum.uva.nl/rob/Courses/InformationInSpeech/CDROM/Literature/LOTwinterschool2006/szabo.best.vwh.net/smart.contracts.html.
5. N. Szabo, "Formalizing and securing relationships on public networks," First Monday, 1997, doi: 10.5210/fm.v2i9.548.
6. A. R. Hevner, S. T. March, J. Park, and S. Ram, "Design science in information systems research," MIS Quarterly: Management Information Systems, 2004, doi: 10.2307/25148625.
7. A. R. Hevner, "A Three Cycle View of Design Science Research," Scandinavian Journal of Information Systems, 2007, doi: http://aisel.aisnet.org/sjis/vol19/iss2/4.
8. M. Schäffer, M. di Angelo, and G. Salzer, "Performance and Scalability of Private Ethereum Blockchains," 2019, doi: 10.1007/978-3-030-30429-4_8.
9. J. Benet, "IPFS - Content Addressed, Versioned, P2P File System," Jul. 2014, [Online]. Available: http://arxiv.org/abs/1407.3561.
10. M. Scherer, "Performance and Scalability of Blockchain Networks and Smart Contracts," White Paper, 2017.
11. PING, "Ping to Integrate Chainlink Oracles to Provide IoT Data to Smart Contracts," 2020, [Online]. Available: https://medium.com/@pingnet/ping-to-integrate-chainlink-oracles-to-provide-iot-data-to-smart-contracts-2fd9d5a1abe2.